\documentclass[prl,twocolumn]{revtex4-1}
\usepackage{latexsym, color, graphicx, comment}
\usepackage{amssymb}
\usepackage{amsmath}

\newcommand{\vect}[1]{\mbox{\boldmath $#1$}}

\newcommand{\energy}{\mathcal{E}}

\newcommand{\changed}[1]{{#1}}
\newcommand{\gstwo}{{\tt gs2}}
\newcommand{\gene}{{\tt gene}}

\begin{document}

\title{Universal instability for wavelengths below the ion Larmor scale}



\author{Matt Landreman}
\email[]{mattland@umd.edu}
\author{Thomas M. Antonsen, Jr}
\author{William Dorland}
\affiliation{Institute for Research in Electronics and Applied Physics, University of Maryland, College Park, MD, 20742, USA}


\date{\today}

\begin{abstract}

We demonstrate that the universal mode driven by the density gradient
in a plasma slab
can be absolutely unstable even in the presence of reasonable magnetic shear.
Previous studies from the 1970s that reached the opposite conclusion
used an eigenmode equation limited to $L_x \gg \rho_i$, where $L_x$ is the scale
length of the mode in the radial direction, and $\rho_i$ is the ion Larmor radius.
Here we instead use a gyrokinetic approach which
does not have this same limitation.
Instability is found for perpendicular wavenumbers $k_y$ in the range $0.7 \lesssim k_y \rho_i \lesssim 100$,
and for sufficiently weak magnetic shear: $L_s / L_n \gtrsim 17$, where
$L_s$ and $L_n$ are the scale lengths
of magnetic shear and density.
Thus, the plasma drift wave in a sheared
magnetic field may be unstable even with no temperature gradients, no trapped particles, and no magnetic curvature.

\end{abstract}


\maketitle 

\section{Introduction}

Following a series of papers in 1978 \cite{Ross, Tsang, Antonsen},
it has been widely accepted that
drift waves in a plasma in a sheared magnetic field, in the absence of parallel current, temperature gradients,
and magnetic curvature, are absolutely stable.
In other words, the ``universal mode'' driven by the density gradient in slab geometry\cite{Galeev,Krall3}
becomes absolutely stable if any magnetic shear is included.
\changed{
This conclusion is important
as a basic problem of plasma physics, and for understanding
the physical mechanisms underlying instabilities in more complicated configurations
in laboratory or space plasmas.
For example, the slab limit can provide insight into}
the edge of magnetically confined plasmas,
in which the slab-like drive from radial equilibrium gradients ($\sim 1/L_\perp$ for scale length $L_\perp$)
exceeds the
curvature drive ($\sim 1/R$ for major radius $R$),
and in which the temperature gradient may tend to be relatively weaker than the density gradient
since the former is more strongly equilibrated over ion orbits \cite{Grisha1}.
However, in contrast to the 1978 work
(and later studies \cite{LCN,Chen}),
here we demonstrate that the universal mode is in fact unstable
for a range of perpendicular wavenumbers $k_y$ satisfying $k_y \rho_i \gtrsim 0.7$,
where $\rho_i$ is the ion gyroradius.
We reach a different conclusion to the 1978 work
because these earlier calculations relied on an eigenmode equation derived under the
assumption
$L_x \gg \rho_i$, where $L_x$ is the radial scale length of the fluctuating mode in untwisted coordinates.
In contrast, our gyrokinetic approach does not share this limitation.

\section{Gyrokinetic model}
\label{sec:model}

In slab geometry with magnetic shear, the magnetic field in Cartesian
coordinates $(\hat{x},\hat{y},\hat{z})$ is $\vect{B} = [\vect{e}_{\hat{z}} + (\hat{x} / L_s) \vect{e}_{\hat{y}} ]B$
for some constant $B$ and shear length $L_s$.  A field-aligned coordinate system
is introduced: $x = \hat{x}, \;y = \hat{y} - \hat{x} \hat{z} / L_s, \; z = \hat{z}$,
so $\vect{B}\cdot\nabla x = \vect{B}\cdot\nabla y = 0$ and $\vect{B}\cdot\nabla z = B$.
We consider electrostatic fluctuations satisfying the standard gyrokinetic orderings \cite{RutherfordFrieman, TaylorHastie, Catto}:
 the gyroradii are of the same formal order as the perpendicular wavelengths of the fluctuations, and these scales are
one order smaller than the scale lengths of
the equilibrium density and magnetic field.
We therefore may use the linear electrostatic
gyrokinetic and quasineutrality equations derived in
\cite{RutherfordFrieman, TaylorHastie, Catto},
dropping toroidal effects, temperature gradients, and collisions
\changed{(except where noted in figure \ref{fig:Tnu}.)}
For perturbations varying in $y$ as $\exp(i k_y y)$
and independent of $x$,
the gyrokinetic equation is
\begin{equation}
\frac{\partial h_s}{\partial t}
+ v_{||} \frac{\partial h_s}{\partial z}
=
\frac{q_s}{T_s} f_{Ms}
J_{0s}
\frac{\partial \Phi}{\partial t}
-\frac{i k_y f_{Ms}}{B L_n} J_{0s} \Phi
\label{eq:gke}
\end{equation}
and the quasineutrality condition is
\begin{equation}
\sum_s \left[ -\frac{q_s^2 n_s \Phi}{T_s} + q_s\int d^3v J_{0s} h_s\right]
=0.
\label{eq:qn}
\end{equation}
Here $s \in \left\{i,e\right\}$ denotes species,
$\Phi(t,z)$ is the electrostatic potential, $h_s(t,z,\energy,\mu)$ is the nonadiabatic distribution function,
$\energy = v^2/2$, $\mu = v_{\bot}^2/(2B)$,
and $f_{Ms}$ is the leading-order Maxwellian.
The scale length of the equilibrium electron density $n=n_e$ is $L_n = -n/(dn/dx)$,
$q_i = e = -q_e$ is the proton charge,
$T_s$ is the species temperature,
$J_{0s}(z) = J_0(k_{\perp} v_{\perp} / \Omega_s)$,
$k_{\perp}(z) = k_y \sqrt{1 + (z/L_s)^2}$,
and $\Omega_s = q_s B/m_s$, $J_0$ is a Bessel function.
As $k_y \rho_i$ was ordered $\sim 1$,
(\ref{eq:gke})-(\ref{eq:qn})
are valid for both $k_y \rho_i \ge 1$ and $\le 1$;
the same is true of $k_y \rho_e$
where $\rho_e$ is the electron gyroradius.

For eigenmodes $\propto \exp(-i \omega t)$,
a dispersion equation can be obtained by introducing $\hat\Phi(x)$ satisfying
\begin{equation}
\Phi(z)=
\int dx \; \hat\Phi(x) \exp\left(\frac{i k_y x z}{L_s}\right),
\label{eq:transform}
\end{equation}
with an analogous transform for $h_s$.
We use $x=\hat x$ as the transform variable because
$\hat\Phi(x)$ represents the radial mode in the untwisted coordinates.
Using (\ref{eq:transform}), $\partial /\partial z$ in (\ref{eq:gke})
becomes $i k_{||} = k_y x/L_s$, so (\ref{eq:gke}) may be solved for $\hat h_s$.
Using (\ref{eq:qn}) and approximating $J_{0e}\approx 1$, we obtain \cite{Chen}
\begin{eqnarray}
\label{eq:integral}
0&=&
\left[
\frac{T_e}{T_i}+1
+\left(1-\frac{\omega_*}{\omega}\right)
\left|\frac{L_s}{k_y x}\right|\frac{\omega}{v_e}
Z\left(
\left|\frac{L_s}{k_y x}\right|\frac{\omega}{v_e}
\right)
\right]\hat\Phi(x)
\nonumber \\
&&+\left(\frac{T_e}{T_i}+\frac{\omega_*}{\omega}\right)
\left(\frac{k_y}{2\pi L_s}\right)^2
\int dx' \int dx'' \int dz \int dz'
\nonumber \\
&&\times \Gamma_{0i}(z,z')
\hat\Phi(x'')
\exp\left(\frac{i k_y}{L_s}\left[
x'z-x'z'-xz+x''z'\right]\right)
\nonumber \\
&&
\times
\left|\frac{L_s}{k_y x'}\right|\frac{\omega}{v_i}
Z\left(
\left|\frac{L_s}{k_y x'}\right|\frac{\omega}{v_i}
\right)
\end{eqnarray}
where
\begin{eqnarray}
\label{eq:Gamma}
\Gamma_{0i}(z,z')
&=&I_0\left( \frac{k_y^2\rho_i^2}{2}\sqrt{1+\frac{z^2}{L_s^2}}\sqrt{1+\frac{z'^2}{L_s^2}}
\right)
\\
&&\times \exp\left(-\frac{k_y^2\rho_i^2}{4}
\left[2+\frac{z^2+z'^2}{L_s^2}\right]\right).
\nonumber
\end{eqnarray}
Here, $v_s=\sqrt{2T_s/m_s}$, $\rho_i=v_i/\Omega_i$,
$I_0$ is a modified Bessel function,
$Z$ is the plasma dispersion function,
and $\omega_* = k_y T_e/(e B L_n)$ is the drift frequency.
The multiple Fourier transforms in (\ref{eq:integral}) arise
because the $Z$ function arguments are naturally written in terms of $k_{||} \propto x$,
whereas the Bessel function arguments are naturally written in terms of $z$.
Notice the integral equation (\ref{eq:integral})
from this gyrokinetic approach
is manifestly different from the differential eigenmode equations
in Refs. \cite{Ross, Tsang, Antonsen}.

\section{Numerical demonstration of instability}
\label{sec:numerical}

\begin{figure}[h!]
\includegraphics[width=3.5in]{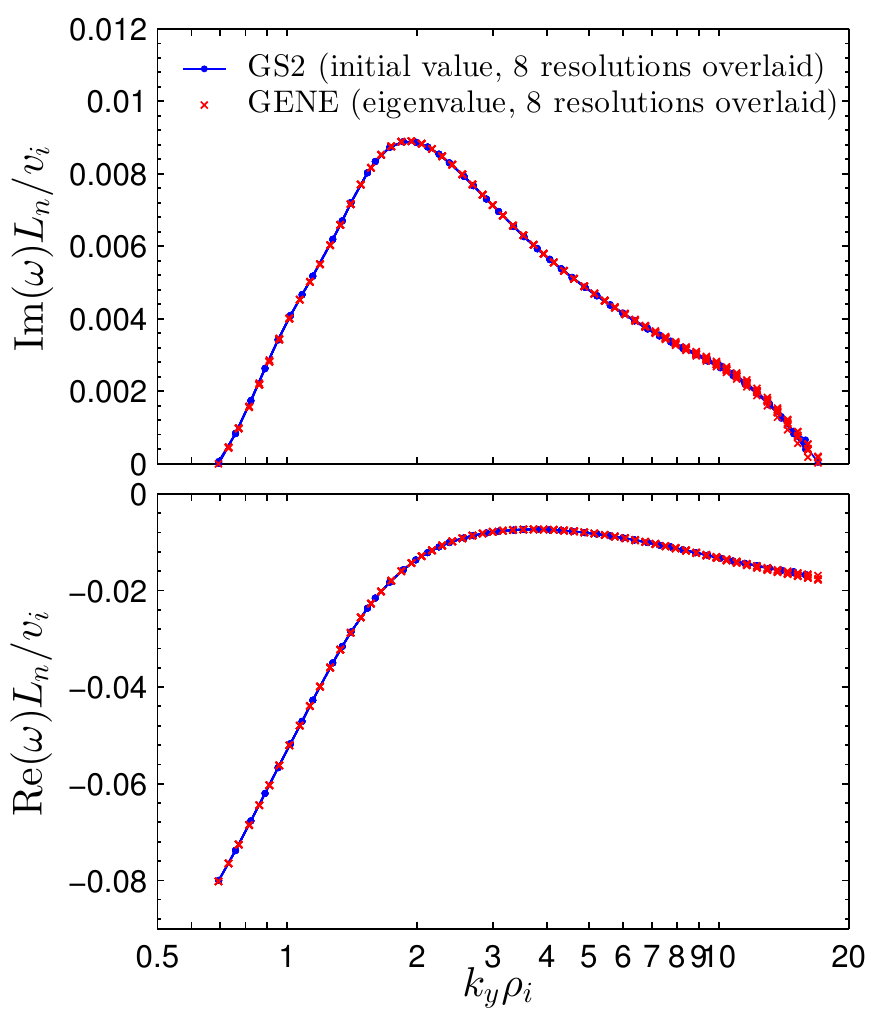}
\caption{(Color online) Demonstration that the universal instability
is numerically robust.
The positive growth rate and real frequency are plotted for the unstable range of $k_y$
using calculations from two independent codes -- \gstwo~run as an initial-value simulation
and \gene~run as a direct eigenvalue solver --
with results agreeing to high precision.
For each code, results from many different numerical resolutions are overlaid
in the figure.
A reduced mass ratio $m_i/m_e=25$ is used here to further minimize the possibility
of numerical artifacts.
\label{fig:convergence}}
\end{figure}

We solve (\ref{eq:gke})-(\ref{eq:qn}) using two independent codes, \gstwo~\cite{gs2} and \gene~\cite{GENE1, GENE2, GENE3}.  A demonstration of linear instability is given in figure \ref{fig:convergence} for $L_s/L_n=32$.
In this figure and hereafter, $T_i = T_e$.
Due to the literature asserting stability,
we have taken great care to verify that this instability
is robust and not a numerical artifact.
To this end, we
demonstrate in figure \ref{fig:convergence}
precise agreement between the two codes, which use different velocity-space coordinates and
different numerical algorithms.
The figure shows calculations from \gstwo~run as initial-value simulations,
and from \gene~run as a direct eigenmode solver using the {\tt SLEPc} library \cite{slepc}.
For each code, results are plotted for many different resolutions to demonstrate convergence.
(For \gstwo, results are overlaid for 8 sets of numerical parameters:
a base resolution, $2\times$ number of parallel grid points, $3\times$ parallel box size, $2\times$ number of energy
grid points, $2\times$ maximum energy, $2\times$ number of pitch angle grid points, $1/2$ timestep, and $2\times$ maximum time.
For \gene, results are again overlaid for 8 sets of numerical parameters:
a base resolution, $2\times$ number of parallel grid points, $2\times$ parallel box size,
$2\times$ number of $v_{\perp}$ grid points, $2\times$ maximum perpendicular energy, $2\times$ number of $v_{||}$ grid points,
$2\times$ maximum parallel energy, and $2\times$ numerical hyperviscosity.)
To further minimize the possibility of numerical artifacts, for this figure
a reduced mass ratio $m_i/m_e=25$ was employed.
As the differences between the 16 series
displayed in each plot are nearly invisible,
the instability is clearly physical and not numerical.

\section{Mode properties}
\label{sec:properties}

\begin{figure}[h!]
\includegraphics[width=3.5in]{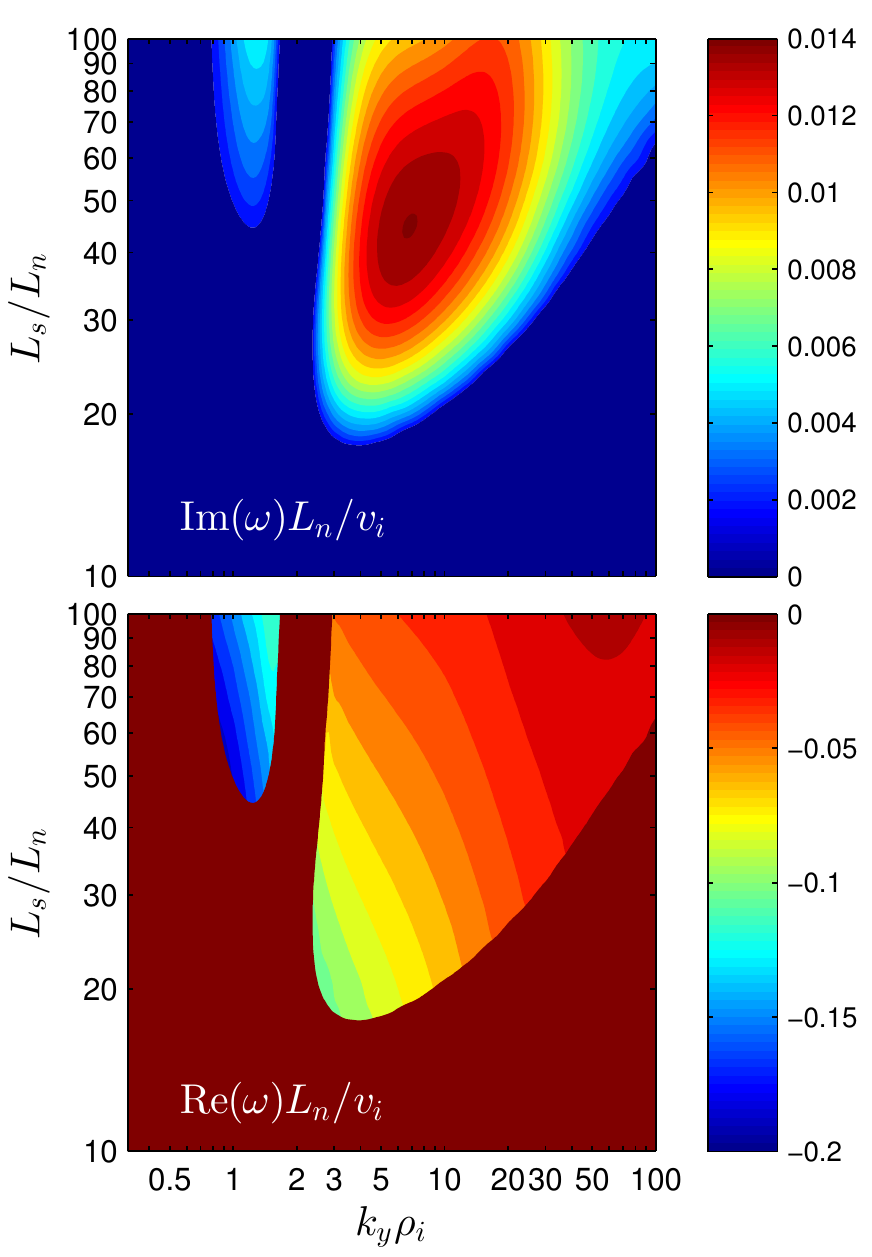}
\caption{(Color online)
Dependence of the universal instability on wavenumber and magnetic shear,
for $m_i/m_e=3600$.
Calculations are performed with the \gstwo~code.
\label{fig:landscape}}
\end{figure}

Figure \ref{fig:landscape} displays the dependence of the frequency and growth rate
on perpendicular wavenumber and magnetic shear, this time using the true deuterium-electron mass ratio.
As one expects from the zero-shear universal instability,
the phase velocity direction is $\mathrm{Re}(\omega/\omega_*)>0$
(appearing in the figures as $\mathrm{Re}(\omega)<0$ due to the codes' sign convention).
The frequency satisfies
$|\omega L_n/v_i | \ll 1$ and $|\omega / \omega_*| \ll 1$.  The growth rate is smaller
in magnitude than the real frequency.

When $L_s/L_n \gg 1$, instability is found for $k_y \rho_i > 0.7$, extending beyond $k_y \rho_i > 100$ for the
highest values of $L_s/L_n$.
Stability for $k_y\to \infty$ can be understood from the fact that $J_{0s}$
in (\ref{eq:gke})-(\ref{eq:qn}) becomes small, as the gyro-motion averages out very small-scale fluctuations.
Stability is the opposite limit of small $k_y$ is expected since Refs. \cite{Tsang,Ross}
are valid there.
A region of stability also exists around $k_y \rho_i \approx 2$ when a realistic mass ratio is used.
No such notch in the growth rate is found when the shear is exactly zero.
This abrupt change in eigenvalue when a small change is made to the physical system
-- in this case the addition of small magnetic shear for $k_y \rho_i \approx 2$ --
is a phenomenon seen in many systems with non-orthogonal eigenmodes \cite{TrefethenEmbree} such as this one.
For such ``non-normal'' systems, the behavior over a finite time is typically a much less sensitive
function of parameters than the eigenvalues, which represent behavior for $t\to \infty$
when all points along the field line have had time to communicate.
Indeed, in initial-value computations we find no notch near $k_y \rho_i \approx 2$ in the growth
for short times $t \lesssim L_s/v_i$, before ions have have traveled far enough in $z$ to ``realize'' there is shear;
these results will be presented in a separate publication.

Instability is found for at least some values of $k_y$ whenever $L_s / L_n$ exceeds a critical
value of $\sim 17$.
Note that in a tokamak, $L_s/L_n \approx q R/(\hat{s} L_n) \gg 1$
can exceed this threshold value.
Here $q$ is the safety factor, $R$ is
the major radius, $\hat{s} = (r/q) dq/dr$, and $r$ is the minor radius.
The general trend in figure \ref{fig:landscape} of decreasing growth rate with increasing shear
can be understood from the observation that in the zero-shear limit,
instability is found when $|k_{||}|$ is below a $k_{\perp}$-dependent threshold.
When sufficient shear is included, all of these unstable parallel wavelengths
are sufficiently long for there to be a significant increase in $k_\perp = k_y \sqrt{1+(z/L_s)^2}$,
causing $\Phi$ to be averaged out by gyroaveraging as discussed above.
For some $k_y \rho_i$ such as $\sim 7$,
the growth rate does not increase monotonically with $L_s/L_n$.
This effect is another example of eigenvalues being sensitive and perhaps misleading,
typical in non-normal systems \cite{TrefethenEmbree} such as this one,
for we find the short-time amplification to increase with $L_s/L_n$ monotonically.

\begin{figure}[h!]
\includegraphics[width=3.5in]{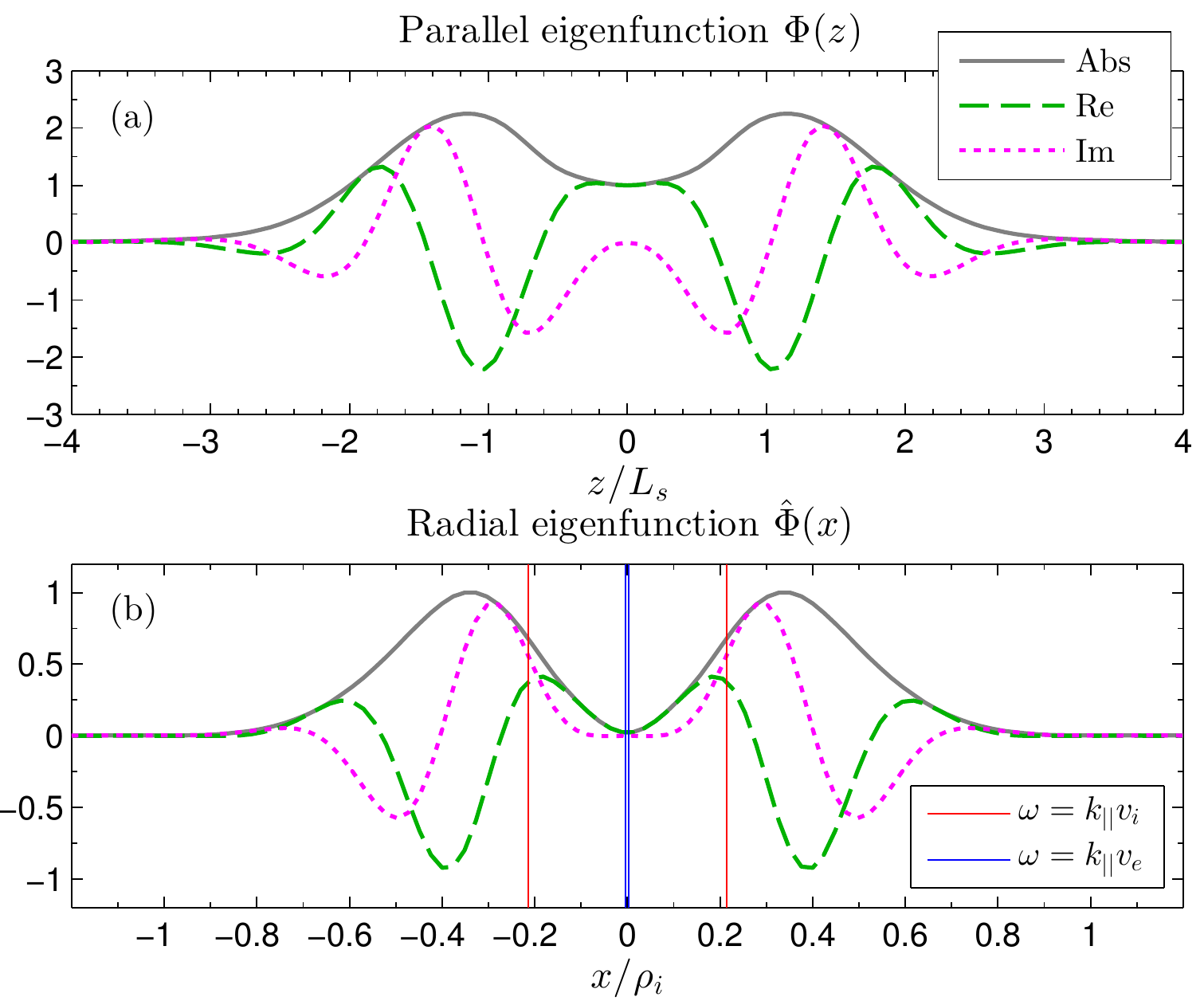}
\caption{(Color online)
Eigenmode computed with \gstwo~for $L_s/L_n=50$, $m_i/m_e=3600$, and $k_y\rho_i=10$
(near the maximum growth rate in figure \ref{fig:landscape}).
In (b), noting $k_{||}=k_y x/L_s$,
vertical lines indicate the values of $x$ at which the arguments of the electron
and ion $Z$ functions in (\ref{eq:integral}) have unit magnitude.
\label{fig:eigenmodes}}
\end{figure}

A typical unstable eigenmode is shown in figure \ref{fig:eigenmodes}.
This example is obtained using \gstwo~at $L_s/L_n = 50$ and $k_y \rho_i = 10$, near
the maximum growth rate in figure \ref{fig:landscape}.
In the field-aligned parallel coordinate $z$,
the mode has a parallel extent $\sim L_s$.
Fourier transforming,
the extent of the radial eigenmode is $\lesssim \rho_i$.
Most of the mode structure occurs where the ion $Z$ function
has an argument of order 1,
$|x| \sim |\omega L_s / (k_y v_i)|$,
so the electron $Z$ function has small argument.
The eigenmode amplitude is small in the inner electron region
where $|x| < |\omega L_s / (k_y v_e)|$.
\changed{
The electrons are highly nonadiabatic:
noting the ratio of electron nonadiabatic/adiabatic terms
in (\ref{eq:integral}) is $\approx -(i\sqrt{\pi}/2) (L_s/L_n)
(\rho_e/\rho_i) (\rho_i/|x|)$, then the nonadiabatic electron term
exceeds the adiabatic term wherever $|x|<0.7 \rho_i$,
everywhere the eigenfunction is significant.
}

\begin{figure}[h!]
\includegraphics[width=3.5in]{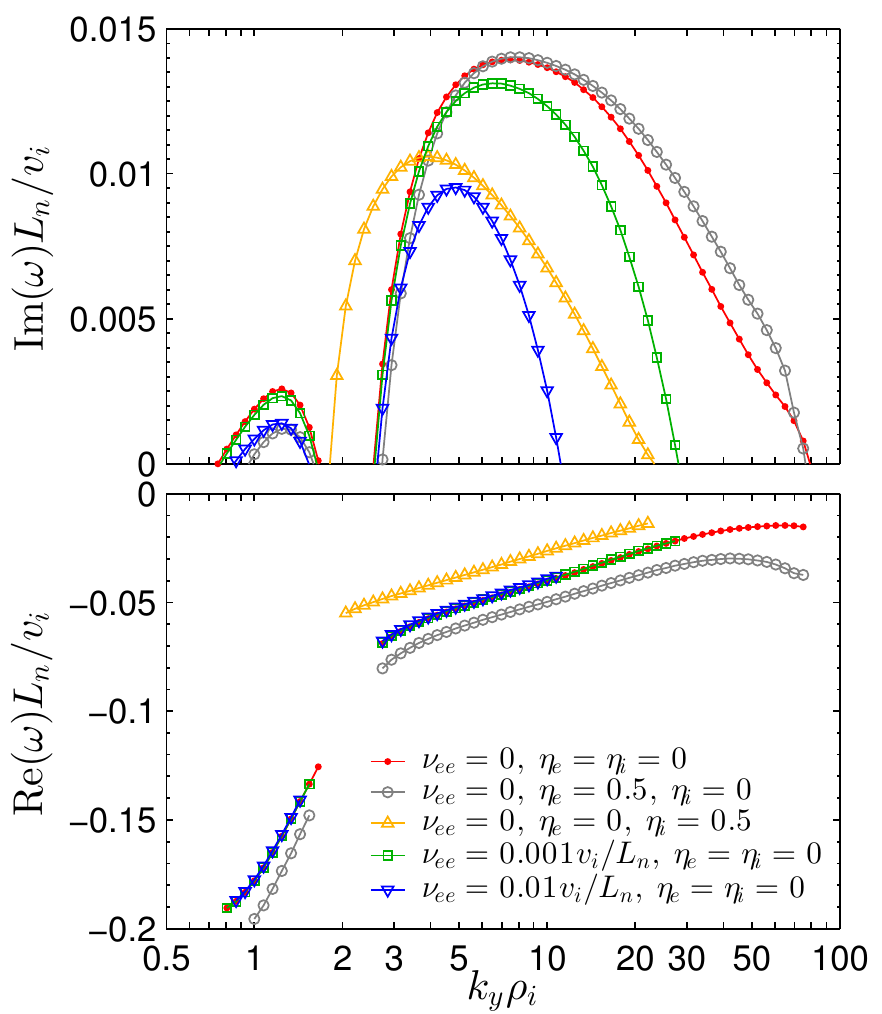}
\caption{(Color online)
Growth rate and real frequency vs. wavenumber for $L_s/L_n=50$ and $m_i/m_e=3600$, showing trends with
$\eta_s = (d \ln T_s/dx)/(d \ln n_s/dx)$ and collisionality.
\label{fig:Tnu}}
\end{figure}

Dependence of the instability on temperature gradients and collisionality is shown in figure \ref{fig:Tnu},
for the case $L_s/L_n=50$ and $m_i/m_e=3600$.
Collisions (using the operator in \cite{Abel}) reduce the growth rate, particularly at the highest $k_y$ values.
This behavior is expected because
(i)
in the free energy moment of (\ref{eq:gke}) (Eq (222) of \cite{AbelReview})
collisions provide an energy sink;
(ii)
the gyrokinetic collision operator \cite{Abel} contains terms $\propto k_\perp^2$; and
(iii) the modes at highest $k_y$ have fine velocity-space structure
(as high resolution in energy and pitch-angle is required for numerical convergence) which
collisions destroy.
For sufficient collisionality, $\nu_{ee} \sim 0.05 v_i/L_n$ or more, the mode is stabilized
at all $k_y$ (for this $L_s/L_n$).

\section{Relationship to previous work}
\label{sec:previous}

Let us now examine in greater detail
why we find instability whereas
previous authors found stability.
First, consider Ref. \cite{Antonsen}, which gave an analytical proof of stability in the limit $T_i \ll T_e$.
We find no contradiction using gyrokinetic simulations, obtaining instability only when $T_i / T_e$ is not small.
(Recall $T_i = T_e$ for all figures here.)

Next, consider Refs. \cite{Ross,Tsang}. These authors considered a differential
eigenvalue equation, not equivalent to (\ref{eq:integral}),
derived assuming
\begin{equation}
\rho_i^2 | \hat\Phi^{-1} \partial^2 \hat\Phi/\partial x^2| \ll 1
\label{eq:differentiallimit}
\end{equation}
(see e.g. \cite{RosenbluthCatto}.)
No such assumption was made in deriving (\ref{eq:integral}),
which is therefore more general.
Considering the typical eigenmode in figure \ref{fig:eigenmodes}.b,
there is structure on scales smaller than $\rho_i$,
violating (\ref{eq:differentiallimit}).
Therefore the approach used in Refs. \cite{Ross,Tsang}
is inapplicable for the modes that are unstable.

Ref. \cite{LCN} claims to give a proof of stability which does
not rely on (\ref{eq:differentiallimit}).  However,
this reference uses an integral eigenmode equation which
differs from (\ref{eq:integral}),
so it is not surprising that we reach different conclusions.
The eigenmode equation in \cite{LCN} is effectively a WKB approximation,
the zero-shear dispersion relation but with $k_x$ replaced by $-i\,d/dx$. This WKB approximation is
not justified for modes such as the one in figure \ref{fig:eigenmodes}, in which the wavelengths in $x$ are comparable to
or much longer than the scale of variation in the ion and electron $Z$ functions.
The problems with the integral equation in \cite{LCN} are
discussed in the introduction of \cite{Chen}
and references therein.

Ref \cite{Chen} appears to give a proof of stability using the
full gyrokinetic integral eigenmode equation (\ref{eq:integral}).
Although we retain $J_{0e}$ for all numerical results shown here while $J_{0e}$ is set to
1 in Ref. \cite{Chen}, this difference is unimportant:
we find little change in the instability for $k_y \rho_i < 20$ if $J_{0e}$ is set to 1 in \gstwo.
The proof in \cite{Chen} is accomplished by multiplying the electron nonadiabatic term
in (\ref{eq:integral}) ($\propto 1-\omega_*/\omega$)
by a parameter $\lambda$.
The authors first prove stability in the adiabatic electron limit $\lambda=0$.
We do not disagree,
as we find no instability numerically with adiabatic electrons.
However we do disagree with the final step
of the proof, where it is argued that solutions
for $\omega$ cannot cross the real axis as $\lambda$ is increased from 0 to 1 to recover (\ref{eq:integral}).
Including $\lambda$ in \gstwo, we find that modes do continuously transform
from damped to unstable as $\lambda$ is increased from 0 to 1. For the case
$L_s/L_n=50$ and $k_y \rho_i=10$ shown in figure \ref{fig:eigenmodes},
marginal stability occurs when $\lambda=0.032$.
Therefore,
the problematic step in \cite{Chen} is item (iii) on p750.
Indeed, when $\lambda>0$, a nonadiabatic electron term must be included
in (38)-(55), and this term can have opposite sign to the other terms in (55),
so (56) no longer follows.

\section{Conclusions}
\label{sec:conclusions}

In summary, we find the plasma slab with weak or moderate magnetic shear ($L_s/L_n > 17$) is generally unstable at
low collisionality, even when temperature gradients are weak.
The density-gradient-driven electron drift wave,
known to be absolutely unstable in the absence of magnetic shear \cite{Galeev, Krall3},
is not stabilized at all wavelengths when small magnetic shear is introduced,
counter to previous findings.
This instability is seen robustly in multiple gyrokinetic codes,
and occurs for $k_y \rho_i \gtrsim 1$.
Previous work that apparently showed stability of the universal mode \cite{Ross, Tsang, Antonsen}
assumed either (\ref{eq:differentiallimit}) or $T_i \ll T_e$,
whereas we make neither assumption.
Ref. \cite{LCN} employed a less accurate dispersion relation, and
\cite{Chen} appears incorrect for nonadiabatic electrons.
\changed{
Though it has sometimes been assumed that an electrostatic instability seen in a gyrokinetic simulation
propagating in the electron diamagnetic direction must be a trapped electron mode or
electron temperature gradient mode,
our results indicate neither trapped particles nor temperature gradients
are necessary for instability in this phase-velocity direction.}
As the universal mode can indeed be absolutely unstable in the presence of magnetic shear,
it should be considered alongside temperature-gradient-driven
modes, trapped particle instabilities, ballooning modes, and tearing modes
as one of the fundamental plasma microinstabilities.

\begin{acknowledgments}
This material is based upon work supported by the
U.S. Department of Energy, Office of Science, Office of Fusion Energy Science,
under Award Numbers DEFG0293ER54197 and DEFC0208ER54964.
Computations were performed on the Edison system at
the National Energy Research Scientific Computing Center, a DOE Office of Science User Facility supported by the Office of Science of the U.S. Department of Energy under Contract No. DE-AC02-05CH11231.
We wish to thank Tobias G\"{o}rler for providing the \gene~code.
We also acknowledge conversations about this work
with Peter Catto, Greg Hammett, Adil Hassam, Edmund Highcock,
Wrick Sengupta, Jason TenBarge, and George Wilkie.
\end{acknowledgments}

\appendix


%

\end{document}